\documentclass[journal=jacsat,manuscript=article]{achemso}
\usepackage{float}
\usepackage{subfig}
\usepackage{natbib}
\usepackage[version=3]{mhchem} 
\usepackage{xr}
\usepackage{amsmath}
\usepackage{amssymb} 
\usepackage[numbers]{}

\makeatletter
\newcommand*{\addFileDependency}[1]{
  \typeout{(#1)}
  \@addtofilelist{#1}
  \IfFileExists{#1}{}{\typeout{No file #1.}}
}
\makeatother

\newcommand*{\myexternaldocument}[1]{%
    \externaldocument{#1}%
    \addFileDependency{#1.tex}%
    \addFileDependency{#1.aux}%
}

\myexternaldocument{SUPPLEMENTARY/supplementary}

\usepackage{hyperref}
\hypersetup{
    colorlinks=true,
    linkcolor=blue,
    filecolor=magenta,      
    urlcolor=red,
    citecolor=blue,
    }

\author{Alireza Soleimani}
\altaffiliation{Department of Physics, Technical University Dortmund, Dortmund 44221, Germany}
\author{Herre Jelger Risselada}
\altaffiliation{Department of Physics, Technical University Dortmund, Dortmund 44221, Germany}
\affiliation[Goettingen University]
{Institute for Theoretical Physics, Georg-August-University Göttingen, Göttingen 37077, Germany}
\alsoaffiliation{Faculty of Science, Leiden Institute of Chemistry, Leiden University, Einsteinweg 55, 2333CC, Leiden, The Netherlands}
\email{jelger.risselada@tu-dortmund.de}

\title[An \textsf{achemso} demo]
  {SMARTINI3: Systematic Parametrization of Realistic Multi-Scale Membrane Models via Unsupervised Learning and Multi-Objective Evolutionary Algorithms}
  
\begin{document}
\begin{abstract}

\noindent \textbf{Abstract} \\
\noindent In this study, we utilize genetic algorithms to develop a realistic implicit solvent ultra-coarse-grained (PC) membrane model comprising only three interaction sites. The key philosophy of the ultra-CG membrane model SMARTINI3 is its compatibility with realistic membrane proteins, for example, modeled within the Martini coarse-grained (CG) model, as well as with the widely used GROMACS software for molecular simulations. Our objective is to parameterize this ultra-CG model to accurately reproduce the experimentally observed structural and thermodynamic properties of PC membranes in real units, including properties such as area per lipid, area compressibility, bending modulus, line tension, phase transition temperature, density profile, and radial distribution function. In our example, we specifically focus on the properties of a POPC membrane, although the developed membrane model could be perceived as a generic model of lipid membranes. To optimize the performance of the model (the fitness), we conduct a series of evolutionary runs with diverse random initial population sizes (ranging from 96 to 384). We demonstrate that the ultra-CG membrane model we developed exhibits authentic lipid membrane behaviors, encompassing self-assembly into bilayers, vesicle formation, membrane fusion, and gel phase formation. Moreover, we demonstrate compatibility with the Martini coarse-grained model by successfully reproducing the behavior of a transmembrane domain embedded within a lipid bilayer. This facilitates the simulation of realistic membrane proteins within an ultra-CG bilayer membrane, enhancing the accuracy and applicability of our model in biophysical studies.
\end{abstract}







\section{1. Introduction}

Lipid bilayer membranes are believed to be a vital component of living cell machinery\cite{sadeghi2018particle}. In addition to their role in various biochemical functions such as facilitating synergy among different lipids, cellular signaling and cellular transport, lipid bilayers also function as barriers, separating cells and organelles from their external environment \cite{martinez2013molecular, moradi2019correction, sadeghi2021hydrodynamic}. Currently, multiple computational models are utilized to simulate biomembranes across different scales \cite{lindahl2000mesoscopic,marrink2001effect,de2005molecular,moradi2019shedding,gu2020phase,goetz1998computer,drouffe1991computer,noguchi2001self,wang2005modeling,farago2003water,cooke2005tunable,cooke2005solvent,mori2016molecular,arnarez2015dry,sharma2021evaluating,liu2019molecular}. Specifically, coarse-grained (CG) models, where a single interaction site represents a group of particles, have garnered considerable attention in the field in recent years because of significant alleviation of the spatial and temporal limitations of atomistic molecular simulations\cite{Why_CG}. Pushing the degree of coarse-graining to an extreme, certain models have been developed to represent an entire bilayer section using one single interaction site (e.g., Refs. \cite{yuan2010one, fu2017lennard, feng2018entropic}). Clearly, such a simplified model does not preserve the lateral nature of biological bilayers, which involves the distribution of material across two distinct, adjoined leaflets and therefore the occurrence of so-called lipid flip-flops between them \cite{allhusen2017ins}. The bilayered structure of biological membranes plays a crucial role in various significant biological processes, including pore formation \cite{awasthi2016simulations, tieleman2006lipids}, membrane curvature formation \cite{bubnis2016exploiting}, membrane fusion and fission \cite{smirnova2010solvent, mattila2015hemi}, and notably in lipid droplet formation, where the bilayer separates into two monolayers to encapsulate a hydrophobic droplet of neutral lipids \cite{zoni2019bud}. To maintain the bilayered nature of biological membranes, several ultra-coarse-grained models have been suggested, representing a lipid molecule with only three interaction sites. These models employ either tabulated potentials derived through systematic coarse-graining approaches \cite{srivastava2013hybrid} or adopt an explicit analytical form for the coarse-grained potential \cite{cooke2005solvent, PhysRevE.72.011506, PhysRevE.72.011915, xu2015mesoscale}. In these phenomenological models, the lipid topology is typically depicted either as a linear chain comprising two (or more) hydrophobic tails beads and one hydrophilic tail bead (e.g., Refs. \cite{cooke2005solvent, PhysRevE.72.011506, PhysRevE.72.011915}) or as a head group bead connected to two individual hydrophobic tails (e.g., Refs. \cite{xu2015mesoscale}). There exists a pressing necessity to advance the development of ultra-coarse-grained (ultra-CG) models, aiming to facilitate highly efficient simulations of more complex and realistic biological systems and to also go beyond phenomenological levels within ultra-CG models of lipid membranes. In pursuit of this objective, these models should accurately replicate both the structural and thermodynamic properties of lipid membranes. Simultaneously, they should support multi-scale approaches, facilitating the incorporation of membrane proteins modeled at significantly higher, or even atomistic, detail. Systematic coarse-graining approaches utilized in the development of realistic ultra-coarse-grained (ultra-CG) models include force-matching approaches \cite{izvekov2004effective, izvekov2005multiscale} and inverse Boltzmann and inverse Monte Carlo approaches \cite{lyubartsev1995calculation, lyubartsev2005multiscale}. These approaches parameterize coarse-grained force fields by replicating the structural component of the partition function of the fine-grained system by either matching relevant radial distribution functions or (combined) forces within the fine-grained system. However, since the partition function only describes a single thermodynamic state point at equilibrium, i.e., a unique combination of pressure \& temperature values, such a systematic structure-focused parameterization may not inherently translate into an optimal reproduction of elastic properties or accurate description of phase transitions. Furthermore, the model obtained often characterizes a specific system and may not be directly transferable to various systems. Alternatively, ultra-CG models are constructed using a building block approach, i.e., all nonbonded (and bonded potentials) are parametrized by the same predetermined potential function, and distinctions between interaction types are achieved through the tuning of only a few free parameters, for example, $\sigma$ and $\epsilon$ in case of the famous Lennard-Jones potential. This approach enhances transferability, allowing the model to be applied to different systems. It also promotes compatibility with coarse-grained representations of other membrane constituents, including various lipid types and membrane proteins, potentially even at a different model resolution. Regrettably, the 'bottom-up' development of force-fields is a challenging task because the values of free parameters are not straightforwardly derived and demand extensive fine-tuning to optimally match the desired structural and thermodynamic properties. Consequently, models developed in this manner are typically utilized as phenomenological models only. In this study, we employ Multi-objective Evolutionary Algorithms (Genetic Algorithms) to construct a realistic three-site coarse-grained membrane model of POPC that optimally reproduces experimentally known structural properties as well as thermodynamic properties. We demonstrate that this ultra-coarse-grained model enables the realistic behavior of lipid membranes, including self-assembly into bilayers, vesicle formation, membrane fusion, as well as the formation of the gel phase. Furthermore, we highlight the compatibility of our model with the widely utilized Martini coarse-grained model for biomolecular simulations. This compatibility allows for the simulation of realistic membrane proteins embedded in an ultra-coarse-grained membrane environment. Significantly, to broaden the accessibility of ultra-coarse-grained (ultra-CG) models to a wider audience, we provide topology files of the potentials derived in this work in a tabulated form, which are directly compatible with the popular molecular dynamics engine GROMACS. Finally, we argue that the here-presented approach can be straightforwardly extended to more complex lipid membrane systems.


\section{2. Methodology}

\subsection{2.1. Force field}

\begin{figure}[H]
\centering
\hspace{10cm}\includegraphics[width=1\linewidth,height=10cm]{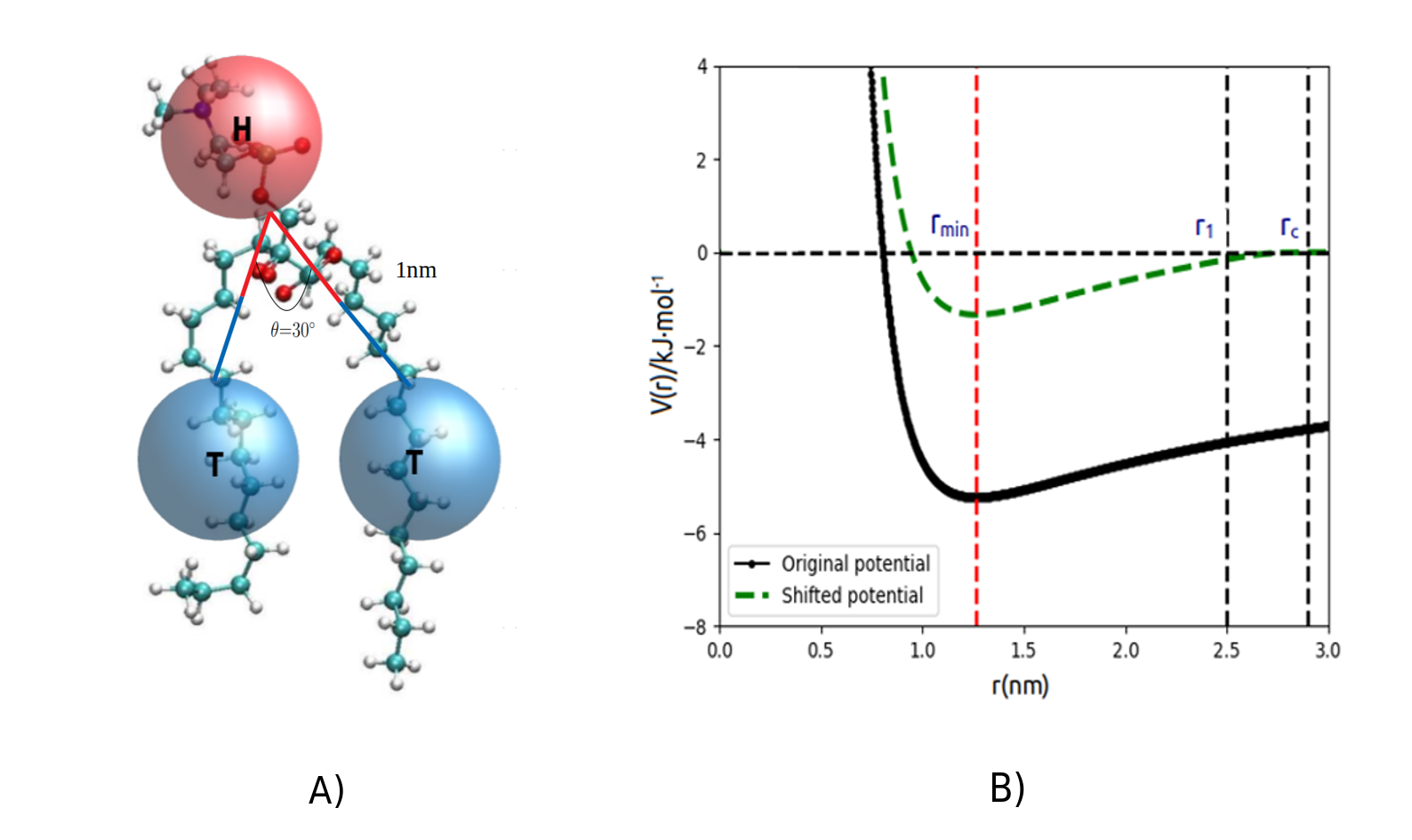}
\renewcommand{\figurename}{Figure}
\caption{A) The schematic drawing depicts the coarse-grained model of a lipid molecule. The hydrophilic bead, represented by the letter "H," corresponds to the cholesterol and phosphate groups of the lipid, whereas the two hydrophobic beads labeled "T" symbolize the fatty acid chains. B) Original and shifted potential functions: Intermolecular potential energy versus distance of a pair of beads. }
\label{fig:lipid_potential}
\end{figure}

In this study, each lipid molecule is characterized by one hydrophilic head and two hydrophobic tails, representing choline and phosphate groups, and the fatty chain of a lipid molecule, respectively. To achieve this representation, we employed a three-site coarse-grained (CG) lipid model, as illustrated in Figure \ref{fig:lipid_potential}A. It is notable that the lipid model utilized in this study is discussed elsewhere in the literature \cite{xu2015mesoscale}. The Hamiltonian of the three-site CG model, considering both bonded and nonbonded interactions, is presented below:

\begin{equation}
U=U_{bond}+U_{angle}+U_{nonbonded}
\end{equation}

\noindent , where $U_{bond}=0.5k_{bond}(r-r_{0})^2$ and $U_{angle}=0.5k_{angle}(\theta-\theta_{0})^2$ denote the harmonic stretching and bending potential, respectively, whereas $U_{nonbonded}$ represents the nonbonded potential for intermolecular interactions. Here, $k_{bond}$ and $k_{angle}$ are the force constants , $r_{0}$ is the equilibrium bond length (1nm) , and $\theta_{0}$ is the equilibrium angle (30$^\circ$). Recent studies in the field \cite{wang2010systematically,ingolfsson2014power, brini2013systematic} have underscored the importance of the attractive potential between lipid tails in implicit solvent models as a crucial factor in replicating the equilibrium structure and self-assembly properties of bilayers. Therefore, the three-site CG model introduced in this study also incorporates an attractive potential between hydrophobic tails. In general, the nonbonded interactions are divided into two components: a repulsive short-ranged interaction, rapidly diminishing with distance, utilized to simulate the excluded volume effect between CG sites, and an attractive long-ranged interaction that incorporates the effective potential of solvent effects. Ultimately, the following combined potential form is employed to characterize the nonbonded interactions among various CG sites:

\begin{equation}
U_{nonbonded}^{TT}=\frac{\epsilon}{5.5}\left[0.5 (\frac{r_{min}}{r})^{6}-6 (\frac{r_{min}}{r})^{0.5}\right]
\end{equation}

\begin{equation}
U_{nonbonded}^{HH/HT}=\frac{0.4\epsilon}{5.5}0.5(\frac{r_{min}}{r})^{6}
\end{equation}

\noindent , where $\epsilon$ represents the potential well depth and $r_{min}$ is the distance of the minimum potential. To mitigate potential artifacts at the cutoff position, the nonbonded potentials are modified using a shift function.

\noindent As a result, both the potential and forces exhibit continuity and a smooth decay to zero between the starting position of the shift function, denoted as $r_{1}$, and the cutoff distance $r_{c}$ (see Figure \ref{fig:lipid_potential}B). The GROMACS shift function used in this work is applied to the force function F(r) and is given by \cite{van2006origin}:

\begin{equation}
S(r,r_{1},r_{c})=\frac{1+A(r-r_{1})^2+B(r-r_{1})^3}{r^{-(\alpha+1)}}
\end{equation}

\noindent , where $\alpha$ is the interaction power which is 0.5 and 6 for attraction and repulsion in our three-site CG model , respectively. In addition, the constants A and B are obtained according to the conditions that the function should be smooth at $r_{1}$ and $r_{c}$ , hence :

\begin{equation}
A=-\frac{(\alpha+4)r_{c}-(\alpha+1)r_{1}}{r_{c}^{\alpha+2}(r_{c}-r_{1})^2}
\end{equation}

\begin{equation}
B=\frac{(\alpha+3)r_{c}-(\alpha+1)r_{1}}{r_{c}^{\alpha+2}(r_{c}-r_{1})^3}
\end{equation}

\noindent Therefore, the total force function is obtained as follows:
\begin{equation}
F_{s}(r,r_{1},r_{c})=\frac{1}{r^{\alpha+1}}+A(r-r_{1})^2+B(r-r_{1})^3
\end{equation}

\subsection{2.2. Molecular dynamics simulations}

Langevin dynamics (LD) was employed for all simulations with a time step of 0.15 ps, utilizing the molecular dynamics (MD) package GROMACS \cite{abraham2015gromacs}. The equation of motion in LD for a system of N particles with masses M is obtained as follows:

\begin{equation}
M\overset{..}{X}=-\nabla U^{LD}-\gamma \overset{.}{X}+\sqrt{2k_{B}\gamma TR(t)}
\end{equation}

\noindent, where U(x) represents the interaction potential of the particles; $\nabla$ is the gradient operator; $\overset{.}{X}$ and $\overset{..}{X}$ denote velocity and acceleration, respectively; $\gamma$ is the friction coefficient, and the third term is the noise term with T as the temperature, $k_{B}$ Boltzmann's constant and R(t) as delta-correlated stationary Gaussian process with zero-mean. Periodic boundary conditions were applied across all three dimensions. Steepest descent algorithm was used to minimize the energy of all configurations, succeeded by a 200 ps molecular dynamics (MD) simulation in the canonical (NVT) ensemble to equilibrate the systems at a temperature of 315 K using the Nose-Hoover thermostat. Pressure coupling was implemented using the Berendsen barostat. It is noted that various bilayer properties, including area per lipid, area compressibility, and bending modulus, were measured by imposing zero surface tension in the xy plane with a compressibility of 1×10$^{-5}$ 1/bar. Simultaneously, pressure was maintained at 1 bar in the z plane with a compressibility of 0. A relaxation time of 1 ps was employed for both temperature and pressure coupling.

\subsection{2.3. Computational analysis of membrane properties}
\subsubsection{2.3.1. Area per lipid}

\noindent Area per lipid is regarded as a key parameter in the characterization of membranes \cite{moradi2019shedding}. The area per lipid plays a crucial role in fine-tuning force fields for membrane systems, making it an indispensable factor. This suggests that accurately calculating structural properties, including the area per lipid, is essential for the molecular dynamics simulation of membranes \cite{patra2003molecular}. Furthermore, the average area per lipid is associated to various membrane properties, encompassing acyl chain ordering, molecular packaging, and area compressibility \cite{frigini2018molecular}. Furthermore, structural dimensions of bilayers, particularly hydrophobic matching and area per lipid, are pivotal parameters in research domains like in vitro studies of integral membrane proteins or peptides, where the frequent reconstitution of membrane lipids is essential \cite{shahlaei2011exploring,shahlaei2011homology,shahlaei2015conformational,kinnun2015elastic,brown2012curvature,botelho2006curvature}. In the context of phospholipid membranes, experimental studies have indicated that the typical range for the area per lipid is reported to be between 0.55 and 0.75 nm$^2$ \cite{sodt2010implicit}. As an example, Kucerka et al. employed hybrid electron density models and determined the area per lipid for POPC, the most abundant lipid in animal cell membranes, to be 0.683 ± 0.015 nm$^2$ \cite{kuvcerka2006structure}. Experimentally determining membrane properties can be challenging. Consequently, atomistic and coarse-grained molecular dynamics simulations have been extensively employed as an alternative, robust, and reliable method for calculating various membrane properties. In molecular dynamics simulations conducted under an NPT ensemble, the dimensions of the simulation box may fluctuate due to pressure coupling. Therefore, in this study, the average area per lipid was determined by calculating the size of the simulation box in the xy plane using the following relation \cite{grote2020optimization}:  

\begin{equation}
A_{Lipid}=\frac{\left\langle L_{x}L_{y} \right\rangle}{N_{lipids}}
\end{equation}

\noindent ,where $L_{x}$ and $L_{y}$ represent the membrane plane simulation box dimensions and $N_{lipids}$ denotes the number of lipids present in one leaflet of the bilayer. Typically, the time evolution of the average area per lipid during the simulation is monitored and visualized graphically. Indeed, fluctuations around an average value are anticipated for the area per lipid. Upon reaching convergence, the average area per lipid is expected to stabilize, showing minimal changes as the system reaches equilibrium. Indeed, Frigini et al. \cite{frigini2018molecular} emphasized that the time-dependent behavior of the area per lipid can serve as a valuable criterion for evaluating whether a membrane system has achieved a steady state. They conducted an investigation into the impact of glyphosate in both neutral (GLYP) and charged (GLYP2) forms on a fully hydrated DPPC lipid bilayer. This was accomplished by observing the time evolution of the area per lipid. Simulation results indicated that, for both charged glyphosate and neutral glyphosate at various G:L ratios, the systems reached stability after approximately 5-10 nanoseconds (ns) of simulation time \cite{frigini2018molecular}. In a comparative study among multiple phospholipids, including POPC including POPC(C$_{42}$H$_{82}$NO$_{8}$P), DOPE(C$_{41}$H$_{78}$NO$_{8}$P), and POPE (C$_{39}$H$_{76}$NO$_{8}$P), Zhang et al. have reported that POPC and POPE phospholipids exhibit the highest (0.683 nm$^2$) and lowest (0.588 nm$^2$) area per lipid, respectively \cite{zhang2022molecular}. Angladon et al. conducted coarse-grained molecular dynamics simulations at 310 K and reported that the area per POPC molecule is 0.676 nm$^2$, which agrees well with the reference value of 0.683 nm$^2$ \cite{angladon2019interaction}. Wang et al. developed an implicit solvent coarse-grained (CG) model for quantitative simulations of POPC bilayers. In the smallest membrane system comprising 72 lipids, equilibrated for about 30 ns in the NP$_{zz}$AT ensemble, the calculated area per lipid was found to be 0.683 nm$^2$. This alignment with the saturated area measured in experiments indicates that the solvent-free CG force field can effectively reproduce the saturated area per lipid obtained from experimental data \cite{wang2010systematically}. Arnarez et al. utilized the Dry Martini force field to replicate various lipid membrane properties, including the area per lipid, bending modulus, and the coexistence of liquid-ordered and disordered domains. For instance, the area per lipid for a POPC bilayer in Dry Martini (0.64 ± 0.1 nm$^{2}$) was smaller in comparison to that of the Wet Martini (0.66 ± 0.1 nm$^{2}$). In contrast, the experimental data ranged from 0.62 to 0.68 nm$^{2}$ and was measured between 297 and 323 K \cite{arnarez2015dry}. Employing a new coarse-grained (CG) model for lipid and surfactant systems, Marrink et al. determined the equilibrium area per lipid for DPPC to be 0.64 nm$^{2}$ at 323 K \cite{marrink2004coarse}. Guo et al. utilized a coarse-grained model for peroxidized phospholipids, built upon the MARTINI lipid force field. They conducted simulations with lipids in a box, considering lateral x and y directions and the transverse z direction, at a temperature of 300 K. As a result, the equilibrium area per lipid for POPC and DOPC phospholipids was measured to be 0.640 ± 0.0008 and 0.669 ± 0.0006 nm$^{2}$, respectively \cite{guo2016peroxidised}. Srivastava et al. determined the area per lipid for highly coarse-grained lipid models of DLPC, DOPC, and mixed DOPC/DOPS to be 0.5763 nm$^{2}$, 0.662 nm$^{2}$, and 0.614 nm$^{2}$, respectively. These values are slightly lower than those calculated in atomistic systems, which are 0.582 nm$^{2}$ and 0.674 nm$^{2}$ for DLPC and DOPC, respectively. This discrepancy is attributed to the softer effective potential energy landscape in coarse-grained systems, leading them to exhibit a smaller area per lipid value than corresponding atomistic systems \cite{srivastava2013hybrid}. Porasso et al. \cite{porasso2012criterion} employed molecular dynamics simulations to investigate the temporal evolution of various lipid properties, including the surface area per lipid for two lipid bilayer systems, namely DPPC (DiPalmitoylPhosphatidylCholine) and DPPS (DiPalmitoylPhosphatidylSerine). The average area per lipid for DPPC was determined to be 0.663 ± 0.008 nm$^{2}$ in the absence of salt at a temperature of 330 K \cite{nagle1996x,rand1989hydration,nagle2000structure}. Furthermore, the area per lipid for the bilayer formed by 288 DPPS in the presence of CaCl$_{2}$ at 0.25N was calculated as 0.522 ± 0.007 nm$^{2}$ \cite{porasso2012criterion}. It is important to highlight that in our evolutionary runs, the goal is to achieve a specific target area per lipid of 0.68 nm$^{2}$, a value commonly reported for the area per lipid of a POPC membrane in both experiments and simulations. Nevertheless, converged solutions within the broader range of 0.55 nm$^{2}$ to 0.77 nm$^{2}$ are also taken into consideration, aligning with experimental data obtained for PC membranes, as discussed earlier.

\subsubsection{2.3.2. Area compressibility modulus}

Area compressibility stands as a crucial elastic characteristic often used to gauge the reliability and efficiency of membrane models. Compressibility, defined as the volume change of a material under stress, applies specifically to the lipid bilayer's resistance to isotropic area dilation, characterized by an area compressibility modulus, K$_{A}$ \cite{moradi2019shedding}. Experimental techniques, such as micropipette aspiration, have reported the area compressibility modulus for POPC bilayers in the range of 208 to 237 mN/m at 21 °C \cite{rawicz2000effect}. Furthermore, alternative experimental values ranging from 180 to 330 mN/m, acquired through Infrared measurements at 25 °C, have been reported for POPC and SDPC in other studies \cite{binder2001effect}. In MD simulations, various methods exist to compute the area compressibility modulus of a bilayer membrane. In this study, area compressibility or the stretching modulus K$_{A}$ is calculated through an area fluctuation analysis according to the following equilibrium statistical relation:

\begin{equation}
k_{A}=\frac{k_{B}T}{4(\left\langle L_{x}^{2} \right\rangle-\left\langle L_{x} \right\rangle^{2})}
\label{mylinetension}
\end{equation}

\noindent , where $\left\langle L_{x} \right\rangle$ and $\left\langle L_{x}^{2} \right\rangle$  denote the mean value and the quadratic fluctuation of the lateral size of the anisotropic simulation box. Guo et al. computed the stretching modulus for Martini coarse-grained POPC and DOPC bilayers, each comprising 512 lipids, to be 379 ± 21 and 357 ± 19 mN/m, respectively. However, these values were observed to overestimate the accepted experimental reference values. Conversely, in larger systems containing 8192 lipids, better agreement was achieved with compressibilities of 245 ± 42 and 230 ± 27 mN/m \cite{guo2016peroxidised}. Additionally, Xu et al. determined the area compressibility modulus of a tensionless coarse-grained bilayer membrane model through the thermal fluctuations of the membrane area, yielding K$_{A}$= 157.5 mN/m, which was smaller than the experimental data for a DPPC bilayer, i.e., K$_{A}$= 231 ± 20 mN/m \cite{xu2015mesoscale}. Similarly, utilizing fluctuations in the membrane area per lipid, Marrink et al. observed that for a large bilayer patch comprising 6400 lipids, the area compressibility was determined to be K$_{A}$= 260 ± 40 mN/m which is lower compared to the area compressibility of K$_{A}$= 400 ± 30 mN/m for a smaller system consisting of 256 lipids. The observed discrepancy was attributed to the contribution of undulatory modes in the large system, leading to a decrease in the area compressibility. The value obtained for the large system aligns well with the experimental value reported for DPPC, which is  K$_{A}$= 231 ± 20 mN/m \cite{marrink2004coarse}. Furthermore, Wang et al. investigated the temperature dependence of the area compressibility by analyzing the standard deviation in the area per lipid. They identified a correlation between the compressibility modulus and the phase of the system. They observed that systems with a compressibility modulus lower than 210 mN/m existed in the liquid phase, whereas those with a compressibility modulus higher than 420 mN/m were in the gel phase \cite{wang2016dppc}. This discovery suggests that the area compressibility can serve as an indicator of membrane phase transitions. Moreover, as an alternative approach, Saeedimasine et al. calculated the area compressibility modulus by determining the derivative of surface tension with respect to areal strain. Consequently, the calculated area compressibility modulus for Pure-POPC was found to be 188 mN/m at the atomistic level. In contrast, at the coarse-grained level, the values varied between 282 and 297 mN/m for different system sizes \cite{saeedimasine2019role}. Finally, it is noteworthy to mention that our evolutionary trials are designed to achieve a specific target area compressibility of 230 mN/m for a POPC membrane. This chosen value falls within the reported range for POPC membrane area compressibility obtained through both experimental and computational techniques. However, we also consider solutions within the broader range of 180 mN/m to 330 mN/m, as discussed earlier, given that these values align with experimental data for PC membranes.

\subsubsection{2.3.3. Line tension}

\begin{figure}[H]
\centering
\hspace{0cm}\includegraphics[width=0.5\linewidth]{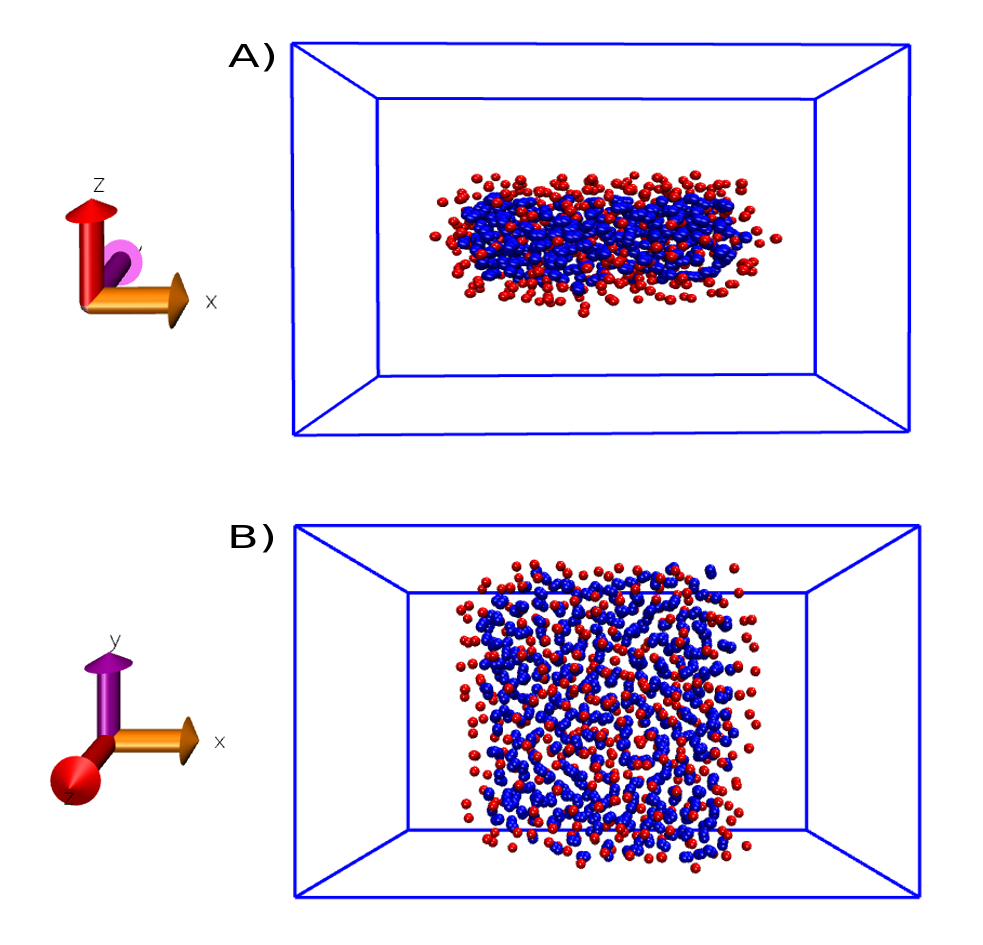}
\caption[Simulation setup for calculation of line tension in a membrane...]{Simulation setup for calculation of line tension in a membrane consisting of 376 lipids. A) side view. B) top view.}
\label{fig:LT_Setup}
\end{figure}

Line tension, alternatively known as edge energy, refers to the excess free energy arising from the existence of free edges \cite{xu2015mesoscale}. The line tension of pores in phospholipid bilayers plays a crucial role in pore-mediated molecular transport techniques. The existence of a pore within the bilayer acts as a gateway for extracellular molecules to enter the cell. This pore facilitates passive molecular transport, enabling molecules to traverse the membrane without requiring additional energy expenditure \cite{shigematsu2015line}. Various methods, encompassing both experimental and computational approaches, have been utilized to measure line tension. In general, the line tension or edge tension in diverse membrane compositions has been experimentally measured and observed to fall within the range of 8 pN to 42 pN \cite{garcia2007pore,may2000molecular,melikyan1990influence,portet2009visualization}. For example, Portet et al. introduced a novel technique that allows for the measurement of edge tensions in lipid membranes. This method involves the electroporation of giant unilamellar vesicles and the subsequent analysis of the dynamics of pore closure. To assess the edge tension, the researchers applied this technique to membranes with various compositions, including dioleoylphosphatidylcholine (DOPC) and mixtures of DOPC with cholesterol. As a result, the line tension of DOPC was obtained to be 27.7 ± 2.5 pN and 36.4 ± 1.9 pN for membranes made of DOPC/cholesterol at a 5:1 molar ratio. Additionally, the reported range of line tensions for a phosphatidylcholine (PC) bilayer membrane typically falls between 6.5 and 30 pN \cite{wang2010systematic}. In this study, molecular dynamics simulations are employed to calculate the line tension  $\gamma_{L}$ from the anisotropy of the pressure within the simulation box. Specifically, this calculation is performed for a bilayer membrane with two free edges in the x-direction, while the membrane is periodically extended in the y-direction (Figure \ref{fig:LT_Setup}) \cite{wohlert2006free}:

\begin{equation}
\gamma_{L}=-\frac{L_{x}L_{z}}{2}(P_{yy}-\frac{P_{xx}+P_{zz}}{2})
\end{equation}
\noindent ,  where $P_{xx}$ and $P_{yy}$ represent the tangential and $P_{zz}$ denotes the normal component of the diagonal pressure tensor, $L_{x}$ and $L_{z}$ are the simulation box dimensions and the prefactor $\frac{1}{2}$ is included to account for the fact that the system has two free edges. Employing a similar methodology, Miyazaki et al. obtained the line tension of a pure ribbon-like POPC bilayer membrane to be 35.1 pN \cite{miyazaki2019free}. Similarly, Y. Jiang et al. utilized pressure anisotropy in the simulation box and calculated the line tension within the range of 10 to 30 pN for the edge of a pure dimyristoylphosphatidylcholine (DMPC) membrane \cite{jiang2004molecular}. Alternatively, employing an implicit solvent coarse-grained (CG) model for POPC bilayers, Wang et al. utilized a straightforward formula, i.e, $\gamma_{L}=-\frac{1}{2}\left\langle \sigma_{xx} \right\rangle L_{y}L_{z}$ to compute the line tension, where $\sigma_{xx}$ represents the xx-component of the stress tensor, $L_{y}$ and $L_{z}$  represent the length of the simulation box in y and z dimensions, respectively. As a result, the calculated line tension value ($\gamma_{L}$= 29 ± 12 pN) fell well in the range of both experimental and atomistic simulation reference values of PC membranes \cite{wang2010systematically}. Given the scarcity of reliable statistical data concerning the experimentally determined value of the POPC membrane line tension, as highlighted in litrature \cite{akimov2017pore}, our evolutionary runs are aimed at achieving a specific target line tension of 10 pN. This value aligns within the experimental range of PC membrane line tension, typically reported between 6.5 pN to 30 pN.

\subsubsection{2.3.4. Bending modulus}

\begin{figure}[H]
\centering
\hspace{0cm}\includegraphics[width=0.8\linewidth,height=10.5cm]{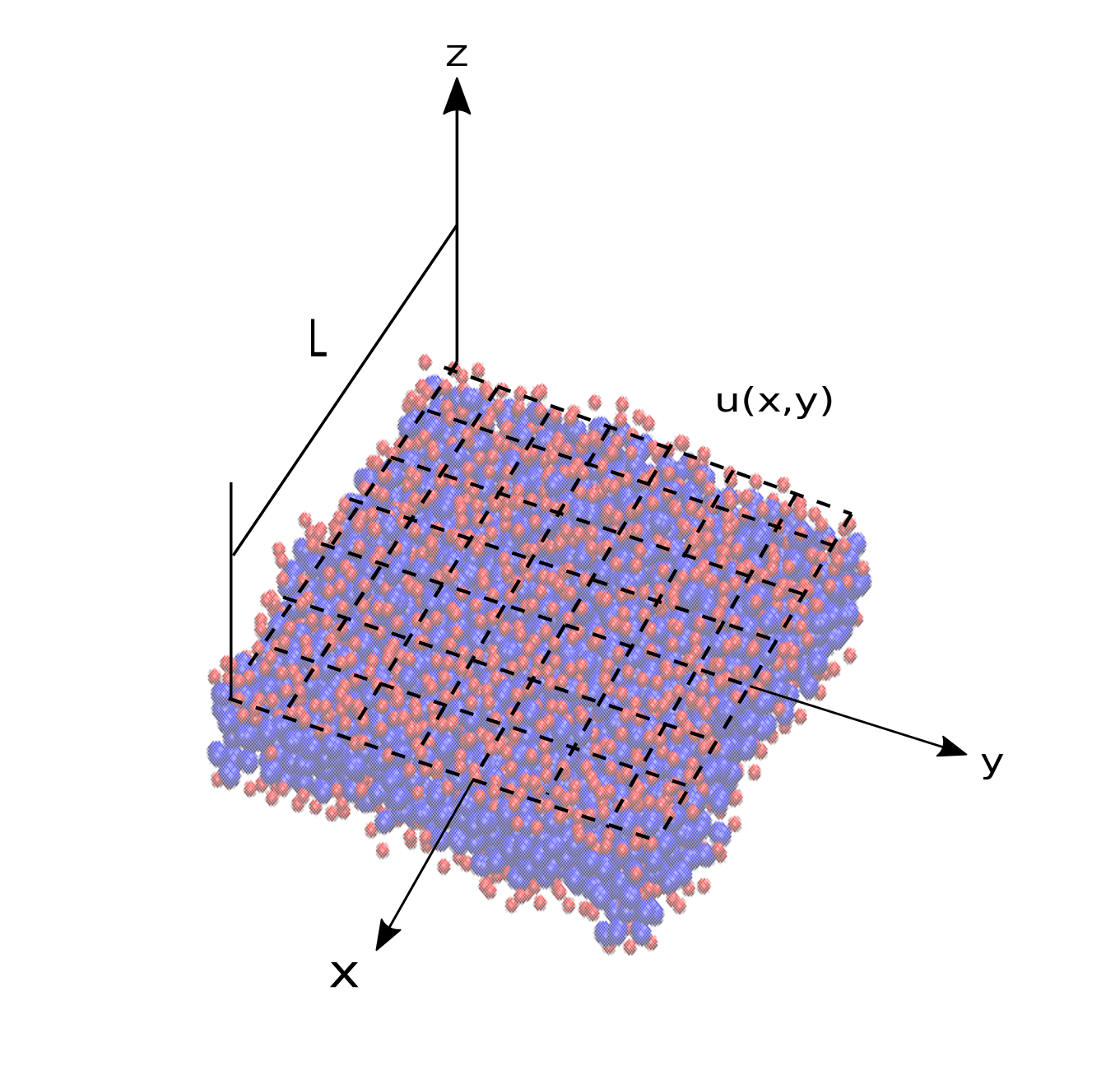}
\caption[Shown is a schematic illustration of the height function in a bilayer membrane...]{The schematic depicts the height function in a bilayer membrane, u(x, y), fitted to bead positions observed in selected frames throughout the simulation trajectory.}
\label{fig:BM_MESH}
\end{figure}

The membrane's bending modulus plays a crucial role in regulating the out-of-plane bending deformation of a membrane as described in the Canham–Helfrich elastic model \cite{guo2016peroxidised}. Bending modulus values were obtained through experimental measurements. Specifically, Rawicz et al. utilized the micropipette vesicle aspiration technique to measure the bending modulus of fluid phase phosphatidylcholine (PC) membranes, i.e, POPC and DOPC, resulting in values of approximately 9.0 $\times 10^{-20}$J and 8.5 $\times 10^{-20}$J, equivalent to around 20 k$_{B}$T and 19 k$_{B}$T, respectively \cite{rawicz2000effect}. In computer simulations, as well as in the present study, the bending modulus for lipid bilayers is commonly determined by computing the membrane undulation spectrum. While the theoretical concept of using the membrane undulation spectrum is straightforward, its practical application is highly sensitive to errors. Moreover, it is only applicable in the "long-wavelength" limit and exhibits slow convergence, as noted elsewhere \cite{brannigan2004solvent}. Utilizing a Helfrich-type continuum model, the bilayer can be represented as a single mathematical surface, denoted as u(x, y), with a minimum energy state. This model facilitates the characterization of the bilayer. Consequently, the Hamiltonian for thermally excited energy fluctuations takes the following form \cite{brandt2011interpretation}:

\begin{equation}
H\{ u(x,y) \}=\frac{1}{2}\int_{}^{}\int_{}^{}(K_{c}|\nabla ^{2}u(x,y)|^{2} + \gamma |\nabla u(x,y)|^{2})dxdy    
\label{eq-Hamiltonian}
\end{equation}

\noindent , where $k_{c}$ and $\gamma$ represent bending modulus and surface tension, respectively.\\
\noindent The membrane fluctuations are analyzed by employing a Fourier decomposition of the surface:

\begin{equation}
u(\textbf r)=\sum_{\textbf q}^{} u(\textbf q)e^{i\textbf q.\textbf r}            
\end{equation}

\noindent , where $r$=($x$,$y$) is a two-dimensional real space vector and $q$=($q_{x}$,$q_{y}$) being the two-dimensional reciprocal space vector. Due to the quadratic nature of energy in the amplitudes (Eq.\ref{eq-Hamiltonian}), the equipartition theorem is applied, stating that each Fourier mode has an average energy of k$_{B}$T/2 \cite{levashov2008equipartition}. In the absence of any in-plane tensions ($\gamma$=0), thermal fluctuations as a function of wave number q are expressed by the following equation \cite{sadeghi2020large}:

\begin{equation}
\langle|u_{q}|^2\rangle=\frac{k_{B}T}{L^{2}k_{c}q^4}
\end{equation}

\noindent ,where $\langle|u_{q}|^2\rangle$ represents the ensemble average of the energy of the undulatory modes, $k_{c}$ denotes the bending modulus , $k_{B}$ is the Boltzmann's constant, T is the temperature and L is the periodic simulation box side. If the membrane model accurately replicates the continuum behavior dictated by the Helfrich energy density, a deviation from the q$^{-4}$ behavior is anticipated at short wavelengths. This divergence arises because the continuum behavior is only applicable to high-wavelength undulations \cite{sadeghi2018particle}. Various methods can be utilized to extract the height spectrum from simulations of bilayer membranes. These methods enable the extraction of the bending modulus, with the Real Space (RS) method being a commonly employed approach for this purpose. In the Real Space (RS) method, the membrane height field, represented as $u(x, y)$, is interpolated onto a uniform grid of nodes in a two-dimensional (N$\times$N) grid (see Figure \ref{fig:BM_MESH}). To mitigate technical complications, such as "empty" grid cells, it is advisable to select a larger value for N in the Real Space (RS) method. Increasing the size of the grid enhances the accuracy of the interpolation and improves the resolution of the height-height spectrum. Moreover, to map the membrane position onto the defined grid, a straightforward approach involves selecting a reference position in each lipid, such as any of the particles or beads belonging to the lipid or the average position of the phosphocholine head group. The height of the reference position for all lipids within a given grid point is then computed, and the average value of the heights is assigned to that grid point. Another approach to mapping lipid molecules onto a uniform grid, without the need for a reference point, involves directly mapping every microscopic degree of freedom. This entails mapping each individual heavy atom or coarse-grained bead to its nearest grid point, irrespective of the lipid to which it belongs. Finally, to acquire the mean power spectrum, the power spectra of all bilayer configurations are averaged. Indeed, numerous publications in the literature have employed the computational approach described above to calculate the bending modulus of membranes. Guo et al. conducted calculations of the bending modulus utilizing a coarse-grained model specifically developed for peroxidized phospholipids, built upon the MARTINI lipid representation. Their findings revealed bending modulus values of approximately 30.0 k${B}$T and 28.3 k${B}$T for POPC and DOPC, respectively \cite{guo2016peroxidised}. Yuan et al. employed a one-particle-thick fluid membrane model, wherein each particle represented a cluster of lipid molecules, to derive the bending rigidity of membranes. Their computations resulted in bending rigidity values falling within the experimental range of 12 k$_{B}$T to 40 k$_{B}$T \cite{yuan2010one}. Srivastava et al. estimated the bending modulus for highly coarse-grained DLPC membranes to be 6.65$\times$10$^{-20}$J, approximately equivalent to 15 k$_{B}$T. Similarly, the bending modulus for a DOPC model was found to be (7.64 ± 0.671)$\times$10$^{-20}$J, corresponding to 18-19 k$_{B}$T \cite{srivastava2013hybrid}. Chacon et al. calculated a bending modulus of 21 k$_{B}$T for POPC membranes at 320 K using coupled undulatory (CU) modes to analyze membrane fluctuations. This approach yielded excellent estimates of the bending modulus in line with proposed theoretical predictions \cite{chacon2015computer}. In a distinct study by Xu et al., a bending modulus value of 4.35$\times$10$^{-20}$ J, roughly equivalent to 10 k$_{B}$T, was reported for a coarse-grained DPPC bilayer \cite{xu2015mesoscale}. In conclusion, our evolutionary runs aim to reach a specific target bending rigidity of 12 k$_{B}$T for a pure POPC membrane, as measured in experiments using techniques like electro-deformation or Fluctuation analysis \cite{niggemann1995bending}. Additionally, solutions converging to experimentally observed bending rigidity values between 8 and 42 k$_{B}$T are considered valid and fall within an acceptable range.

\subsubsection{2.3.5. Phase transition temperature}

\begin{figure}[H]
\centering
\hspace{0cm}\includegraphics[width=0.6\linewidth,height=10cm]{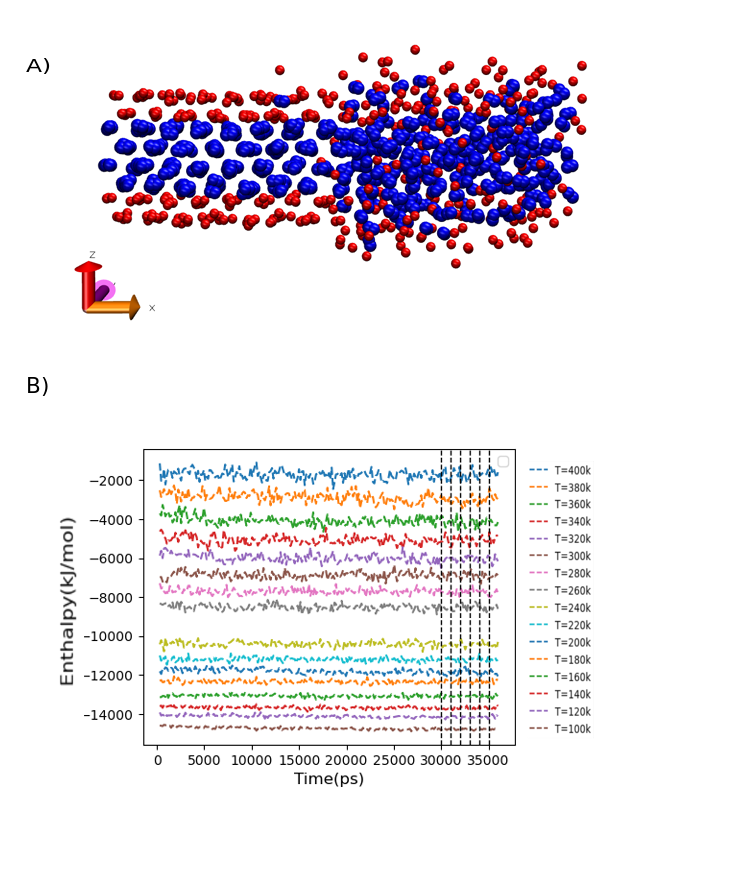}
\renewcommand{\figurename}{Figure}
\caption[Phase transition temperature calculation...]{Phase transition temperature calculation. A: The simulation snapshot depicts the initial configuration of a bilayer, with one half of the lipids in the gel phase (left) and the other half (right) in the fluid phase. 
B: Enthalpy versus time at different temperatures ranging from 100k to 400k. A sudden increase in enthalpy within the temperature range from 240 K to 260 K is indicative of a phase transition from the gel to the liquid-disordered phase. The phase transition temperature is then assigned to the middle of the identified phase transition zone, i.e, 250 k. }
\label{fig:PTT_half}
\end{figure}

Considerable research efforts have been directed towards investigating the phase transition in biological membranes, with a particular focus on the ordered (gel) and disordered (liquid-crystalline) lipid phases. These phases play a crucial role in understanding the overall dynamics and structural properties of membranes, given their direct association with the individual lipid components \cite{leekumjorn2007molecular}. Numerous experimental studies have explored the phase behavior of phospholipids, employing a variety of techniques. Suurkuusk et al. conducted a comprehensive investigation using calorimetric and fluorescent studies on multi-lamellar vesicles of DPPC. In their study, they observed a heating scan of multi-lamellar vesicles that exhibited a phase transition process reminiscent of a first-order transition, a finding subsequently confirmed through fluorescent measurements \cite{suurkuusk1976calorimetric}. Additionally, Davis et al. employed deuterium magnetic resonance to examine the liquid-crystalline, gel phase, and phase transition of multi-lamellar labeled-DPPC vesicles. Their quantitative measurements based on deuterium spectra revealed a sharp decline in moment values as the vesicles were heated, indicating a first-order transition \cite{davis1979deuterium}. In addition to experimental approaches, computational methods have played a significant role in investigating the phase transition in lipid bilayer membranes. For example, the phase transition temperature was determined by observing a sharp rise in enthalpy during the transition from the gel to liquid crystalline states \cite{sun2018membrane}. Furthermore, an effort was made to address the issue of substantial hysteresis, which can pose challenges in accurately determining the phase transition temperature for the liquid-gel transition. To overcome this challenge, simulations were conducted on a system comprising two distinct halves, with one half in the gel phase and the other in the liquid-disordered phase (refer to Figure \ref{fig:PTT_half}). This configuration effectively diminishes or eliminates the nucleation barrier necessary to initiate either the liquid-disordered or gel phase, leading to a considerable reduction in hysteresis \cite{stevens2004coarse}. On the other hand, de Vries et al. conducted an atomistic molecular dynamics study to investigate the phase behavior of lecithin (DPPC) in water. Their simulation results revealed the spontaneous formation of a ripple gel phase when a fully hydrated bilayer was cooled below the phase transition temperature \cite{de2005molecular}. Accordingly, Leekumjorn et al. employed atomistic molecular dynamics simulations to thoroughly examine the structural properties of saturated DPPC and DPPE lipid bilayers in the vicinity of the main phase transition. Despite their similar chemical structures, with the only difference in their choline and ethanolamine groups, DPPC and DPPE exhibited contrasting transformation mechanisms from an ordered (gel) to a disordered (liquid-crystalline) state. Leekumjorn et al. used the area per lipid and bilayer thickness as criteria to determine the first-order phase transition temperature of DPPC and DPPE bilayers through their simulations. According to their findings, the transition temperature was approximately 305 K for DPPC and 320 K for DPPE. These values were compared to experimental measurements reporting transition temperatures of 315 K for DPPC and 337 K for DPPE. Additionally, using Molecular Dynamics with Alchemical Steps, Fathizadeh et al. calculated the phase transition temperature from the gel to fluid phases in pure DPPC bilayers, obtaining values within the range of 303.15 K to 323.15 K \cite{leekumjorn2007molecular}. Coarse-grained approaches have proven effective in determining the phase transition temperature of lipid bilayers. Marrink et al. conducted simulations using a coarse-grained (CG) model to investigate the transformation between a gel and a fluid phase in DPPC bilayers. They employed a cooling technique on bilayer patches and proposed a four-stage reversible process consisting of nucleation, growth, limited growth, and optimization. Extrapolating their findings to a macroscopic bilayer, they obtained a transition temperature of 295 ± 5 K, which was in semi-quantitative agreement with the experimental phase transition temperature value for DPPC (315 K) \cite{marrink2005simulation}. Kociurzynski et al. determined the phase transition temperature of pure DPPC by observing a sharp increase in the area per lipid during the transition from the gel to liquid crystalline states. Their simulation results indicated a transition temperature of approximately 305 K, showing a slight discrepancy of about 10 K compared to the experimental value of 315 K for DPPC. This difference suggested a small deviation between the simulated and experimental transition temperatures \cite{kociurzynski2015phase, marrink2005simulation}. Mirzoev et al. employed a systematic structure-based coarse-graining approach with a grained DMPC lipid model to examine the state point dependence of lipid bilayers. Their simulations covered a temperature range of 200 to 350 K in 25 K intervals. Through the analysis of the area per lipid and lateral diffusion coefficient, they observed a clear transition between 225 and 250 K, signifying a gel-liquid phase transition \cite{mirzoev2014systematic}. Eventually, Wang et al. utilized coarse-grained solvent-free membrane models to explore the phase transformation in DOPC and POPC membranes. Their simulations indicated that the phase transition in a DOPC membrane occurs within the temperature range of 265 to 295 K, while in a POPC membrane, it takes place between 280 and 310 K. The values obtained from the simulations were slightly lower than the experimental transition temperatures reported for DOPC (around 253 K) and POPC (around 270 K) \cite{wang2010systematic, johansson1993phase, litman1991packing}. As a result, our evolutionary runs aim to achieve a phase transition temperature of 270 K for a POPC membrane, aligning with experimental observations. Additionally, we consider solutions converging toward the experimentally observed phase transition temperature for a PC membrane, spanning from 250 K to 315 K.

\subsubsection{2.3.6. Radial distribution functions}

Radial distribution functions (RDF) are crucial in the analysis of biological membrane systems using molecular dynamics (MD) simulations, offering valuable insights into the system's short and medium-range structural information \cite{mu2016radial}. In this work, the Radial Distribution Function (RDF) is defined according to the formulation by Hub et al. \cite{hub2007short}:

\begin{equation}
g(r)=\frac{n(r,r+\Delta r)}{\rho\Delta V_{2D}}
\end{equation}

\noindent ,where $r$ is the distance in the lateral direction, $\Delta$$V_{2D}$=2$\pi$$r$$\Delta$$r$  being the volume of a shell with radius $r$ and thickness $\Delta$$r$ and n($r$,$r$+$\Delta$$r$) is the number of the  sites within this shell. Given the significant level of coarse-graining in this study, a comparison was exclusively carried out using the Earth Mover's Distance (EMD) technique for the tail-tail radial distribution function (RDF). The comparison involved our three-site coarse-grained (CG) model and the tail-tail RDF derived from an atomistic molecular dynamics simulation of a POPC membrane. The EMD method quantifies the minimal cost required to transform one distribution into another \cite{rubner2000earth}. Hence, if the two RDF distributions—the normalized RDF related to the coarse-grained model and the atomistic RDF—are identical, the resulting EMD value will be 0, signifying their complete similarity.

\subsubsection{2.3.7. Density profile}
\noindent The distribution of mass along the z-axis of the membrane is visualized by computing the mass density profile across the bilayer \cite{moradi2019shedding}. The density profile can offer insights into various properties of the membrane, including the distribution of molecules and the presence of structural features \cite{hamedi2020molecular}. In calculating the mass density profile, it is crucial to account for potential fluctuations in the center of mass (COM) of the membrane during the simulation. Hence, in each frame, the coordinates (x, y, z) of all atoms are determined relative to the instantaneous center of mass (COM), with the z-coordinate set to 0. This approach enables an accurate assessment of the distribution of mass along the membrane's z-axis. Similar to the radial distribution function (RDF), we conducted an Earth Mover's Distance (EMD)-based comparison of the mass density profile between the head groups in our three-site coarse-grained (CG) model and the atomistic model of a POPC membrane. In this context, if the normalized density profile of the coarse-grained model perfectly matches that of the atomistic density profile, the Earth Mover's Distance (EMD) value becomes zero.

\newpage

\subsection{2.4. Genetic algorithms (GA)}

Genetic Algorithms (GA) are developed based on genetic laws and Darwin's theory of evolution aimed to enhance the optimization of problems with numerous variables \cite{angibaud2011parameter}. For example, they are utilized in the determination of molecular dockings within protein-DNA complexes or in the generation of new pharmacophore models through electron density topology analysis \cite{becue2008protein, wang2001automatic}. Generally, a Genetic Algorithm (GA) typically consists of five distinct phases, as outlined below:


\subsubsection{2.4.1. Initial population}
Within a Genetic Algorithm (GA), the search space is represented by a population of individuals, where each individual serves as a potential solution to a specific problem. The individuals in a Genetic Algorithm (GA) are encoded as genomes, which are vectors containing a predefined number of parameters. During the initialization phase, each genome is assigned random values within a predetermined range. The random initialization ensures diversity within the initial population, facilitating a broader exploration of the search space. Subsequently, a cost assessment is carried out to evaluate the quality of each genome. This assessment entails comparing the outcomes generated from Molecular Dynamics (MD) simulations, employing the parameters encoded in the genome, with the target values from a reference database. The cost score is subsequently determined by assessing the degree of agreement between the Molecular Dynamics (MD) simulation results and the specified target values. Genomes that do not meet pre-defined criteria or fail to attain a satisfactory cost score are deemed unfit for further consideration. The eliminated unfit genomes are replaced with new genomes in the subsequent generations, maintaining the population size. In Genetic Algorithms (GA), the initial population size is often chosen to be an even number \cite{sakae2019enhanced}.

\subsubsection{2.4.2. Selection}

The selection operator in a Genetic Algorithm (GA) plays a crucial role in exploiting the advantageous traits of the fittest candidate genomes, aiming to improve and enhance them across successive generations. This accelerates the GA's convergence towards a preferred solution for the optimization problem under study \cite{zames1981genetic}. Numerous selection methods have been reported in the literature, among them are roulette wheel selection (RWS), tournament selection (TOS), and linear rank selection (LRS). In this study, roulette wheel selection (RWS) is utilized, given its widespread adoption as a method in Genetic Algorithms (GA) for selecting genomes for the succeeding generation. The roulette wheel selection (RWS) emulates the concept of survival of the fittest. This method efficiently favors genomes with lower cost scores, providing them with a greater probability of being chosen for the next generation \cite{BEHERA2020349}. Specifically, the RWS assigns a probability of selection P$_{i}$ to each genome i based on its cost value. The probability P$_{i}$ for each genome is defined as:  

\begin{equation}
P_{i}=\frac{F_{i}}{\sum_{j=1}^{PopSize}F_{j}} 
\end{equation}

\noindent , where F$_{i}$ corresponds to the cost of the ith genome. Then, a series of N random numbers are generated and compared with the cumulative probability $C_{i}=\sum_{j=1}^{i}P_{i}$ of the population. Ultimately, the fit genome i is chosen and transferred into the new population provided that $C_{i-1}< U(0,1) \le C_{i}$, where $U(0,1)$ is a uniform random number generated between 0 and 1. In RWS, genomes are allocated to consecutive segments on a line, where the size of each genome's segment is proportionate to its fitness level. Following the generation of a random number, the genome to be selected is determined by identifying the segment on the line that corresponds to the generated random number. The described process continues until the desired number of genomes, often referred to as the mating population, is achieved. This method shares similarities with a roulette wheel, as the size of each segment is directly proportional to its fitness.

\subsubsection{2.4.3. Crossover}
\begin{figure}[H]
\centering
\hspace{0cm}\includegraphics[width=0.90\linewidth]{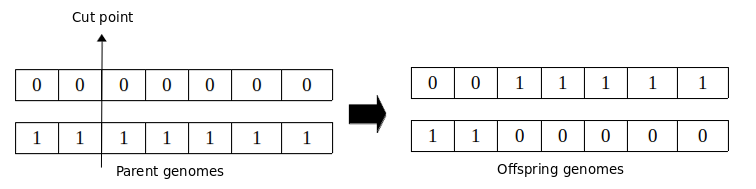}
\renewcommand{\figurename}{Figure}
\caption{Schematic illustration of single-point crossover. The arrow indicates the crossover cut point.}
\label{fig:SPC}
\end{figure}

Undoubtedly, crossover stands out as one of the most crucial operators in Genetic Algorithms (GA), playing a key role in generating offspring that will constitute the subsequent generations. The primary objective of the crossover operation in GA is to broaden the exploration of the search space, thereby aiding in the generation of enhanced genomes in subsequent generations \cite{koohestani2020crossover}. The standard crossover operator involves taking two parent genomes and producing two offspring genomes. The underlying concept behind this operation is that the offspring genomes have the potential to display superior characteristics compared to their parents by inheriting the best traits from each parent. This blending of genetic material from both parents enables the exploration of new genetic combinations that may contribute to improved solutions in the evolutionary process. In this study, single-point crossover is utilized, given its widespread acceptance as one of the most commonly employed crossover operations. In single-point crossover, the two parental genomes are divided into two segments at a randomly determined crossover point. A new child genotype is then generated by exchanging the second segment of the first parent with the second segment of the second parent \cite{kaya2011novel}. Figure \ref{fig:SPC} illustrates the single-point crossover (SPC) operation. \\\\

\newpage

\subsubsection{2.4.4. Mutation}

The mutation operator holds significant importance within a Genetic Algorithm (GA) as it plays a crucial role in preventing premature convergence by introducing a random element in the population's evolution \cite{abdoun2012analyzing}. The mutation operator is applied to the genomes generated through the crossover operation and involves altering one or more bits within a specific genome, with a probability denoted as P$_{m}\epsilon$[0,1), referred to as the mutation probability.  Hence, the mutation probability governs the frequency at which the bits within a particular genome undergo mutation. Significantly, the mutation probability is adjusted based on the behavior of the population. When the population becomes trapped in a local optimum, the mutation probability is increased. Conversely, when the population is widely spread across the solution space, the mutation probability is decreased. However, it is usually chosen within the range of 0.001 to 0.05 \cite{srinivas1994adaptive}. Various mutation operators have been proposed for different solution representations, including Gaussian mutation, random mutation, and polynomial mutation \cite{deb2011multi,rocke2000genetic,deb1999niched}. In this study, Gaussian mutation is employed as the selected mutation operator, given its widespread recognition as one of the most commonly used mutation operators. Accordingly, for each bit in a given genome, a random value is sampled from a normal distribution with a mean of 0 and a standard deviation of 1. Following that, the randomly generated value is compared to a predefined threshold, typically set to 0.05. If the generated value is lower than the threshold, the mutation is considered valid and accepted. However, if the value exceeds the threshold, the mutation is rejected, and no modifications are made to the corresponding bit position within the genome. Once a mutation is accepted, an adaptive mutation operation is executed to introduce additional variability. This operation involves multiplying a sigma value, representing the standard deviation derived from the Gaussian distribution of the bits at a particular position from the previous generation, by the randomly generated number.The resulting value obtained from the adaptive mutation operation is then added to the selected bit within the genome, as illustrated below \cite{bell2022applications}:

\begin{equation}
x_{i} \text{  }\epsilon\text{  }\mathbb{R} \longrightarrow x_{i} \text{  }+\sigma N(0,1)
\end{equation}

\noindent , where i represents the ith bit of genome and $\sigma$ denotes the previously discussed standard deviation.

\subsubsection{2.4.5. Termination}

At this stage of the procedure, the average cost of the best genomes corresponding to a given generation is compared to a threshold, i.e., the target average cost. If the obtained average cost is below the threshold (in the case of a minimization problem) or above the threshold (for a maximization problem), the Genetic Algorithm (GA) terminates. Otherwise, the loop continues until a subsequent generation successfully achieves the goal. It is worth noting that there are alternative stopping criteria for the Genetic Algorithm (GA). These criteria include setting an upper limit on the number of generations or the number of cost function evaluations. Additionally, the Genetic Algorithm (GA) may terminate if the probability of achieving significant changes in the subsequent generation becomes exceedingly low, as mentioned in the referenced book \cite{Book}.

\subsubsection{2.4.6. Cost function}
\noindent  In this study, the weighted sum method, widely recognized as one of the most prominent approaches for addressing multi-objective optimization problems, is employed to calculate the cost of each genome \cite{kim2004adaptive}. This method involves consolidating multiple objectives into a single aggregated objective function by assigning a weighting coefficient to each respective objective function. Hence, the cost corresponding to each genome is calculated using the following equations:\\

\begin{figure}[H]
\centering
\includegraphics[width=0.57\linewidth,height=9.3cm]{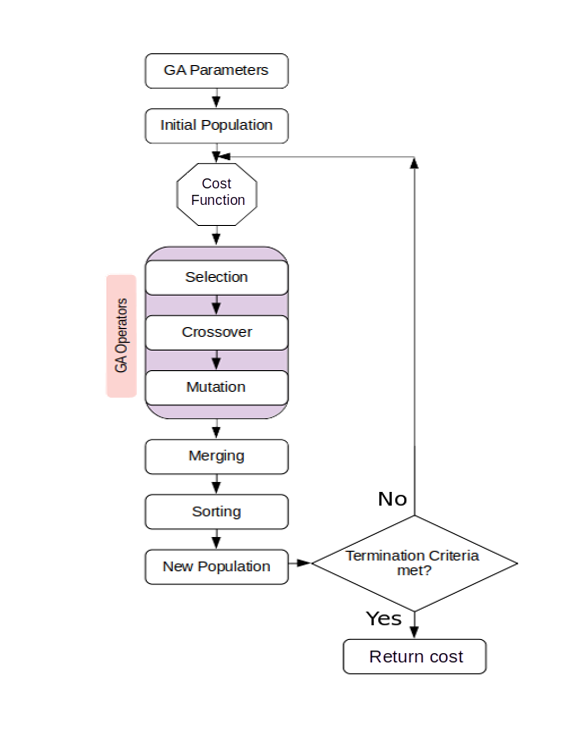}
\caption{Flowchart of the genetic algorithm. }
\label{fig:1}
\end{figure}

\newpage

\begin{equation}
Cost=Cost_{1}+Cost_{2}
\label{equation_total}
\end{equation}
\begin{equation}
\noindent Cost_{1}=\sum_{i=1}^{5}w_{1,i}*[\frac{(Simulated\text{\_}Property)_{i}}{(Target\text{\_}Property)_{i}}-{1}]^{2}
\label{equation_Quantitative}
\end{equation}
\begin{equation}
Cost_{2}=\sum_{i=6}^{7}w_{2,i}*EMD[(CG\text{\_}Simulated\text{\_}Property)_{i},(Atomistic\text{\_}Target\text{\_}Property)_{i}]
\label{equation_Qualitative}
\end{equation}

\noindent ,where $w_{i}$ represents the selected weighting coefficient for the ith objective function. Therefore, equation (\ref{equation_Quantitative}) calculates the cost value for five membrane properties—area per lipid, area compressibility, line tension, bending modulus, and phase transition temperature—each represented by a single value. In contrast, equation (\ref{equation_Qualitative}) is utilized to assess the cost value for the other two properties—radial distribution function and density profile—represented by a distribution. Furthermore, it is important to highlight that the weighting coefficient for each property in the first category is set at 0.15, while for the second category, the assigned weighting coefficient for each property is 0.125. Figure \ref{fig:1} depicts the flow of the genetic algorithm.

\section{3. Results \& discussion}
\begin{figure}[H]
\centering
\hspace{0cm}\includegraphics[width=0.55\linewidth,height=7cm]{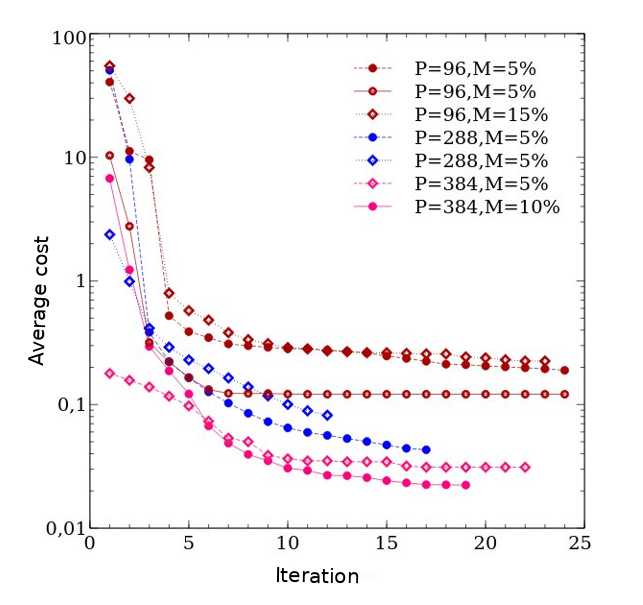}
\renewcommand{\figurename}{Figure}
\caption{Average cost value as a function of Iteration. In the legend, the symbols P and M denote the initial population size and mutation rate, respectively.}
\label{fig:GA_Converge_VESUS}
\end{figure}

\begin{figure}[H]
\centering
\hspace{0cm}\includegraphics[width=0.80\linewidth,height=6cm]{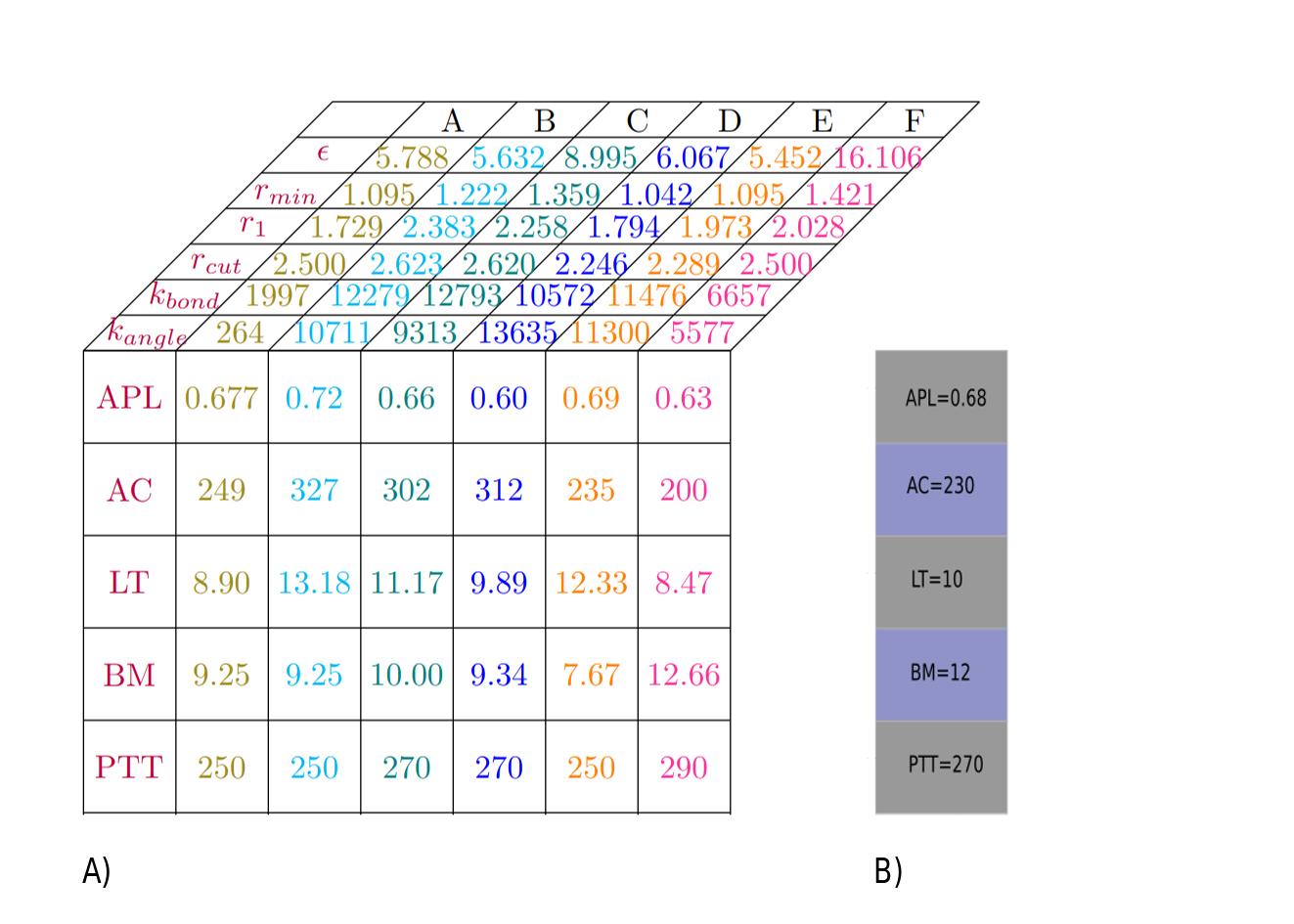}
\caption[A) Illustration of force field parameters as well as the corresponding membrane properties...]{A) Illustration of force field parameters as well as the corresponding membrane properties for 6 different force field parameter sets, each uniquely highlighted by a distinct color. In all depicted 6 sets of force field parameters, the obtained normalized density profile and radial distribution function exhibited complete similarity, i.e, EMD=0, with the reference atomistic density profile and radial distribution function.  B) Predefined GA point targets, i.e, area per lipid (APL= 0.68 nm$^2$), area compressibility (AC= 230 mN/m), line tension (LT= 10pN), bending modulus (BM= 12 k$_{B}$T) and phase transition temperature (PTT= 270K). Alongside the GA point targets, we as well accept genomes that capture membrane properties consistent with experimental data for PC membranes. }
\label{fig:PT}
\end{figure}

\noindent Convergence in genetic algorithms (GA) hinges on various factors, with the initial population size and the number of iterations standing out as two crucial elements. As discussed earlier, the convergence of a genetic algorithm (GA) is significantly influenced by the size of the initial population, as it directly impacts the diversity and exploration capabilities of the algorithm, leading to diverse cost values. Therefore, a larger initial population size tends to facilitate a more extensive exploration of the search space, potentially yielding enhanced genomes (solutions) and cost values that are closer to the desired target. Hence, choosing an appropriate population size is crucial for attaining favorable convergence outcomes in a genetic algorithm (GA). Therefore, in this study, we carried out multiple evolutionary runs, adjusting population sizes from 96 to 384 and mutation rates from 5\% to 15\%. This exploration aimed to determine whether the evolution's convergence resulted in suboptimal (local solution) or optimal (global solution) outcomes. Figure \ref{fig:GA_Converge_VESUS} depicts the average cost versus iteration, demonstrating the evolution of average cost values across successive iterations for each generation. As expected, we observed that increasing the population size led to lower cost values, indicating fitter genomes. For example, when employing an initial population of 96 genomes, the genetic algorithm (GA) converged to a cost value of 0.121. In contrast, with a population size of 384, the GA attained a significantly lower cost value of 0.0242, which is one order of magnitude smaller. This suggests that larger population sizes contribute to smaller cost values in the evolutionary process. Additionally, we recommend a baseline population size of 288 as conducive to achieving convergence in the genetic algorithm (GA). Our simulation results show that commencing with a random population size of 288 enables the genetic algorithm (GA) to meet the desired convergence criteria. Nevertheless, it is crucial to acknowledge that the optimal population size may vary depending on the unique attributes of the problem domain and the characteristics of the search space. Hence, conducting additional experiments and fine-tuning is essential to ascertain the most suitable population size for the genetic algorithm (GA) in a specific problem context. Lastly, Figure \ref{fig:PT} displays six different genomes—force field parameter sets—and their corresponding membrane properties identified by the GA. Each set is represented by a distinct color and is compared against predefined GA targets.


\newpage

\noindent The outcomes of our evolutionary runs reveal that the membrane properties of an optimal genome (solution), identified through the implementation of GA, optimally align with a parameterized force field parameter set including $\epsilon$=5.788 k${_B}$T, r$_{min}$=1.095 nm, r$_{1}$=1.729 nm, r$_{c}$=2.500 nm, k$_{bond}$=1997 kJ.mol$^{-1}$nm$^{2}$, and k$_{angle}$=264 kJ.mol$^{-1}$rad$^{2}$.
Interestingly, a wide range of parameter sets can yield fitness scores that are remarkably close, suggesting that the solution space within these simple models lies on a hyper-spherical plane, with diverse combination of parameters yielding equivalent results. It is crucial to highlight that the parameterization of force-fields often results in an underdetermined system because many of the structural targets characteristic for membranes, such as thickness, area compressibility, line tension, and bending modulus, are interdependent and correlated. This interconnectedness is why coarse-grained models, like the Martini model, can effectively reproduce the elastic properties of lipid membranes without necessitating detailed parameterization of these properties \cite{marrink2004coarse,marrink2007martini}.





\noindent Next, we underscore the significance of the three-site lipid membrane model by examining key processes that demonstrate its ability to accurately mimic the behavior of real lipid membranes \cite{peetla2009biophysical}. These processes encompass the self-assembly of lipids into bilayers, the formation of vesicles through self-assembly, membrane fusion, and the creation of vesicles in the gel phase. These findings highlight the model's utility in biophysical studies, providing insights into the complex dynamics of lipid membranes

\begin{figure}[H]
\centering
\hspace{0cm}\includegraphics[width=0.95\linewidth,height=10cm]{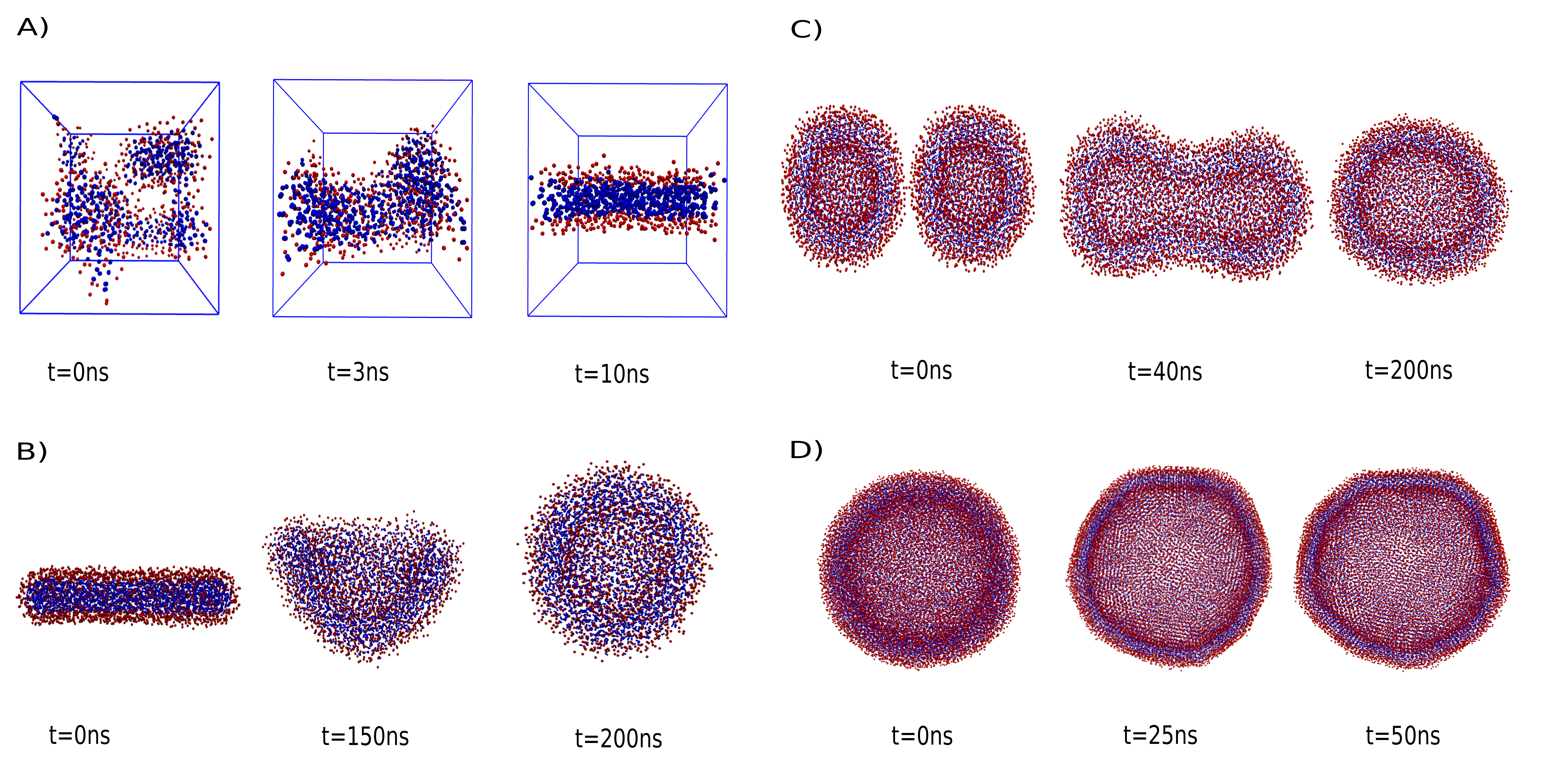}
\renewcommand{\figurename}{Figure}
\caption[Realistic behavior of lipid membranes...]{Realistic behavior of lipid membranes. A) Self assembly into bilayer. B) Self assembly into vesicle. C) Membrane fusion. D) Formation of vesicle in the gel
phase.}
\label{fig:behaviors}
\end{figure}

\noindent A) Self-assembly into bilayers: The fundamental process of lipids self-assembling into bilayers plays a crucial role in the formation of cellular membranes \cite{sych2018lipid}. Lipids, characterized by hydrophilic head groups and hydrophobic tails, spontaneously organize into a bilayer structure, as illustrated in Figure \ref{fig:behaviors}A.

\noindent B) Self-assembly into vesicles: Under specific conditions, lipid membranes can undergo self-assembly, forming spherical vesicular structures. In essence, the self-assembly process results in the spontaneous creation of closed vesicles, featuring a lipid bilayer shell that effectively isolates the internal contents from the external environment. Vesicles hold considerable implications in drug delivery systems, as they possess the ability to encapsulate both hydrophilic and lipophilic drugs. This protective enclosure during transportation not only ensures the integrity of the drugs but also facilitates targeted delivery \cite{xu2018extracellular,burdick1996vesicular,song2022nanoengineering} (refer to Figure \ref{fig:behaviors}B).

\noindent C) Membrane fusion: The dynamic process of membrane fusion occurs when two vesicles make contact and merge, ultimately giving rise to the formation of a continuous vesicle \cite{jahn2003membrane}. This process holds paramount significance in various biological phenomena, encompassing cellular trafficking, exocytosis, and the fusion of enveloped viruses with host cell membranes \cite{rothman1994intracellular,joardar2022mechanism}. Membrane fusion entails a sequence of intricate molecular rearrangements and fusion events that bring two vesicles into close proximity, destabilize their structures, and ultimately result in their merging, as depicted in Figure \ref{fig:behaviors}C.

\noindent D) Formation of vesicles in the gel phase: The existence of lipid membranes in distinct phases is temperature and lipid composition-dependent \cite{heberle2011phase}. In the gel phase, lipids undergo a phase transition, becoming more ordered and rigid. Intriguingly, even in this highly ordered state, lipid membranes retain the capability to form vesicles. The occurrence of vesicle formation in the gel phase is attributed to the local disruption of the lipid bilayer. This disruption enables the bilayer to bend and close upon itself, leading to the formation of vesicles \cite{tallo2021monitoring}. This behavior holds significant implications, particularly in areas like cryopreservation, where the formation of vesicles in the gel phase serves as a protective mechanism for cellular structures against damage induced by freezing (refer to Figure \ref{fig:behaviors}D).

\section{4. Multiscale simulation}

\begin{figure}[H]
\centering
\hspace{0cm}\includegraphics[width=0.95\linewidth,height=14cm]{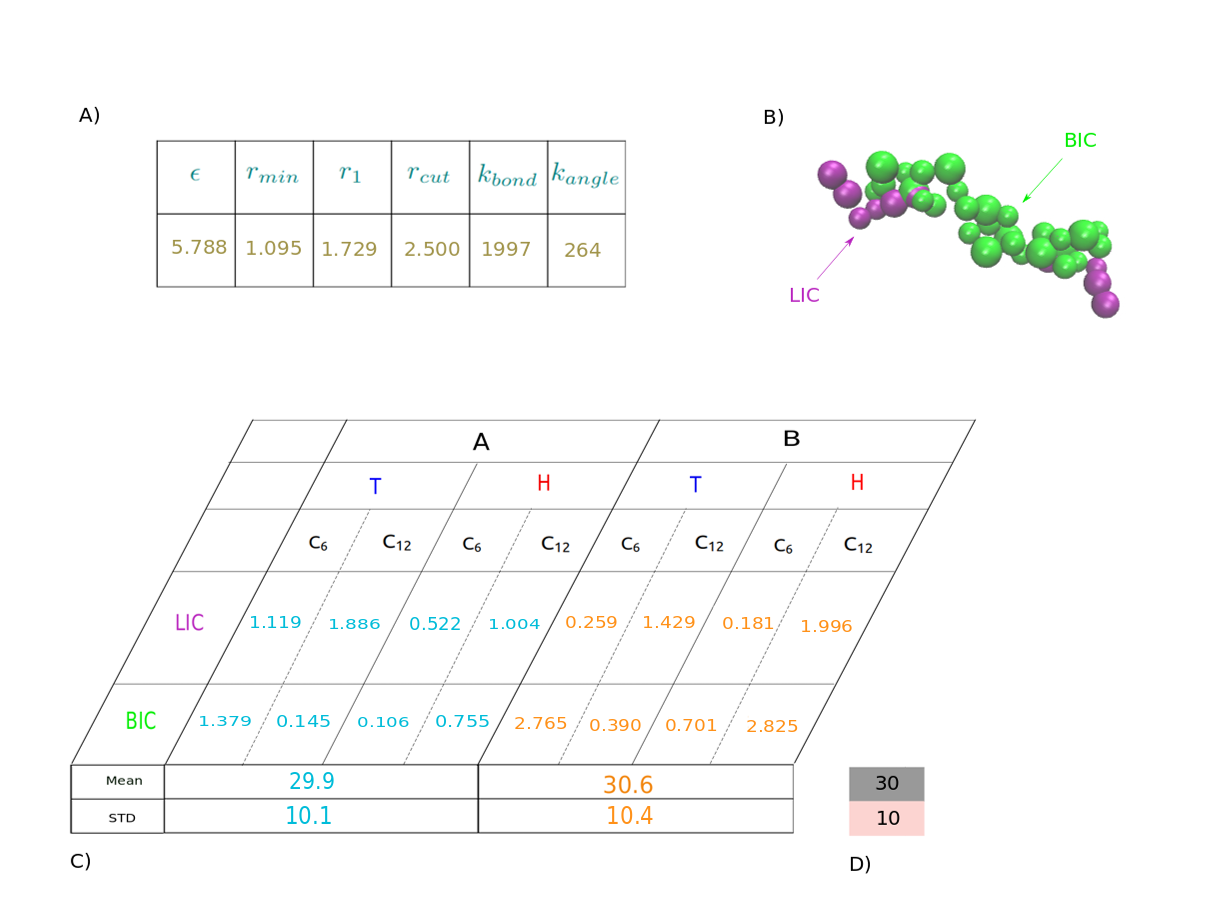}
\renewcommand{\figurename}{Figure}
\caption[A) Membrane force field parameters...]{A) Membrane force field parameters. B) Illustration of Martini KALP Peptide with hydrophilic (LIC) regions represented in purple and hydrophobic (BIC) regions depicted in green. C) Nonbonded membrane-peptide interactions -- C$_{6}$/kJ mol$^{-1}$nm$^{6}$ and C$_{12}$/kJ mol$^{-1}$nm$^{12}$ -- as well as the obtained properties from simulations corresponding to two various force field combinations. D) Predefined GA targets, i.e, probability distribution of the angle defined between peptide and membrane normal with mean and standard deviation of 30$^\circ$ and 10$^\circ$, respectively. }
\label{fig:Multiscale00}
\end{figure}

As highlighted earlier, we believe that it is crucial for the next generations of coarse-grained models to possess state-of-the-art capabilities that facilitate mixed-resolution multi-scale approaches. This ensures seamless integration of membrane proteins, allowing for heightened precision, even at the atomistic level. Therefore, in this study, we introduced an additional technique for conducting multiscale simulations by employing tabulated potentials. These potentials serve as a means to seamlessly integrate various resolution levels. To evaluate the effectiveness of our approach, we conducted a simulation wherein a Martini KALP peptide is embedded in our parameterized coarse-grained (CG) bilayer membrane. The primary goal was to ascertain the stability of the peptide within the membrane. To achieve this, we initially established nonbonded interaction tables to delineate the interactions between various components within the system, as follows:\\
\noindent (i) Interaction tables were generated for the CG membrane beads using a nonbonded cut-off of 2.5 nm, employing the potential forms discussed in Chapter 2.2.1.\\
\noindent (ii) Interaction tables were generated for the Martini peptide beads using a nonbonded cut-off of 1.2 nm, utilizing a modified LJ potential.\\
\noindent (iii) Interaction tables were generated between the Martini peptide and CG membrane beads, utilizing a nonbonded cut-off of 1.2 nm and implementing a modified LJ potential.\\
\noindent Accordingly, the termination criteria for the GA are met when a Gaussian distribution of angles between the bilayer membrane normal and a vector connecting the first bead to the last bead in the peptide is achieved, with a mean of 30$^\circ$ and a standard deviation of 10$^\circ$. Therefore, considering all hydrophilic head groups and hydrophobic tail groups of both the peptide and membrane, the GA needs to parameterize eight nonbonded interactions (C${6}$ and C${12}$) to fulfill its two objectives. In other words, each genome consists of eight parameters (See Figure \ref{fig:Multiscale00}).

\newpage

\begin{figure}[H]
\centering
\hspace{0cm}\includegraphics[width=1.1\linewidth,height=9cm]{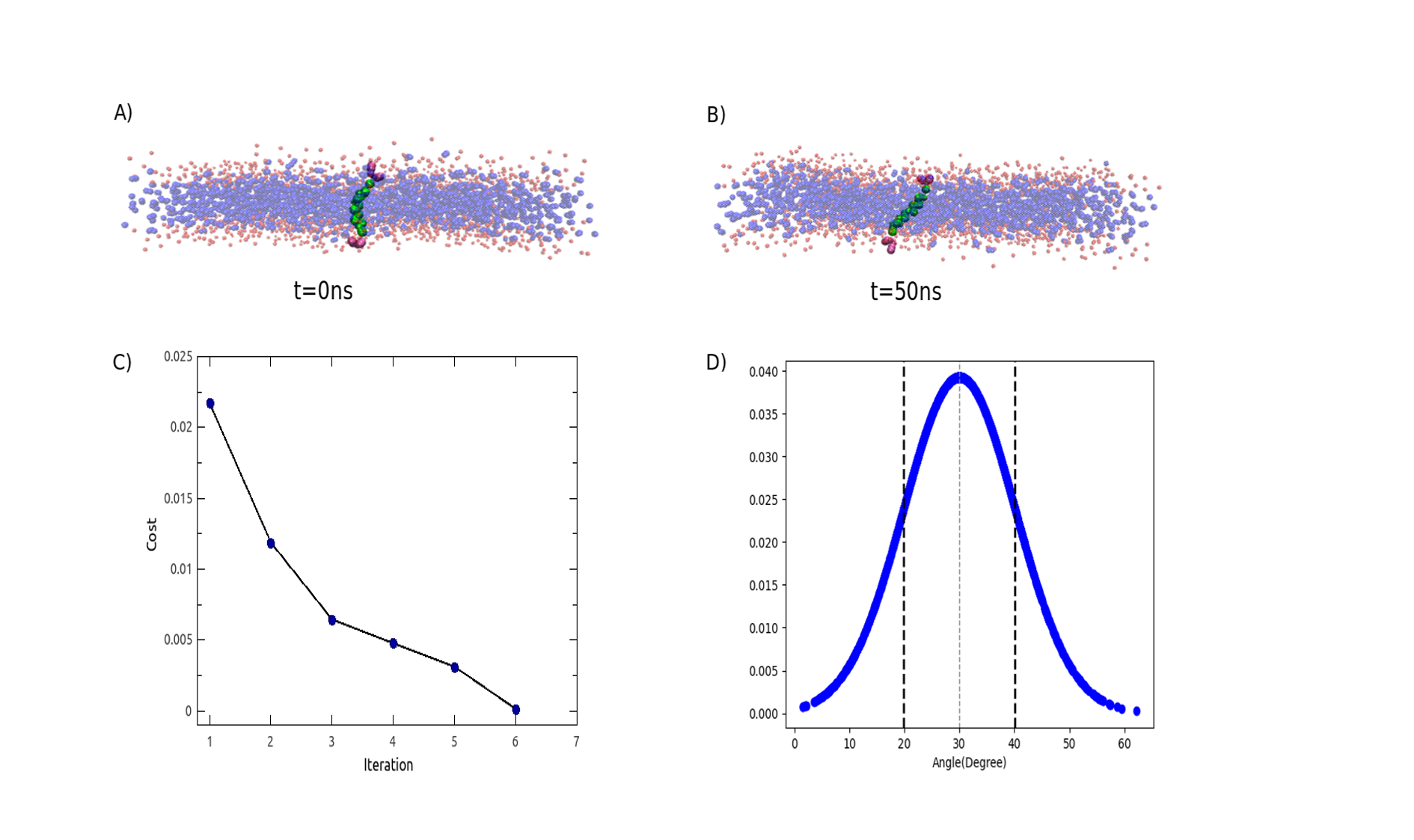}
\renewcommand{\figurename}{Figure}
\caption[Peptide-Membrane Interaction...]{Peptide-Membrane Interaction. A) Snapshot of a KALP peptide within a membrane at the initial of the simulation. B) Simulation snapshot of a KALP peptide within a membrane at t= 50ns utilizing optimized interactions via GA. C) Cost versus iteration. D) GA target, i.e. probability distribution of the angle defined between peptide and membrane normal with mean and standard deviation of 30$^\circ$ and 10$^\circ$, respectively.}
\label{fig:Multiscale}
\end{figure}

\noindent The outcome of the evolutionary process, represented by the cost as a function of iteration, illustrates that the GA successfully achieved the two predetermined objectives within six iterations. Moreover, the angle distribution obtained aligns perfectly with the expected distribution. Furthermore, the simulation snapshots illustrate that despite the initial insertion of the peptide being perpendicular to the membrane, it ultimately adapted to a tilted orientation of 30$^\circ$ due to the parameterization of the genetic algorithm (See Figure \ref{fig:Multiscale}).

Our findings demonstrate a promising synergy between transmembrane peptides and ultra-CG models, achieved through the introduction of only two simple interaction types. A distinguishing feature of our model is its utilization of real units and dimensions, which is crucial for achieving this compatibility. Additionally, the integration of the Gromacs software, which supports the implementation of the Martini model, significantly enhances the model's applicability. 

In our study, we have merely begun to explore the vast potential of hybrid simulation models. We anticipate that such an approach can be straightforwardly extended to more complex membrane proteins, such as viral spike proteins or multi-pass membrane proteins, thereby expanding its utility in biophysical studies. It is crucial to highlight that, given the implicit modeling of solvents, protein-protein interactions outside of the membrane are not realistically represented within a hybrid setup using the standard Martini models. A quick hack to circumvent this issue is to render the protein-protein and protein-membrane interactions within the cytosolic part as being repulsive only. This simple solution could enable the study of scenarios such as the clearance of the membrane fusion site or the expansion of a formed fusion pore, especially when the apposed membranes are overcrowded with sterically repelling proteins that have realistic dimensions, shapes, and membrane behavior. For instance, one could examine the effect of varying protein density on the formation or expansion of a fusion pore, or in the case of enveloped viruses, study the additional release of RNA through a formed fusion pore by combining the model with implicit coarse-grained models for modeling RNA \cite{dawson2016coarse}. Moreover, knowledge of protein-protein interaction site may be ad hoc modeled on a level of non-bonded interactions within Martini, similar to the concept of Martini Go models \cite{poma2017combining}. Certainly, such a setup remains an ultra-CG setup, and as such, the study and system should be adapted to conform with the more limited accuracy of these models. Nonetheless, we anticipate the utility of these models in phenomenologically describing the behavior of large, complex protein-membrane systems with greater accuracy than simplified toy models of proteins.

\section{5. Conclusion}

In this work, evolutionary algorithms, specifically genetic algorithms, are employed to construct an implicit solvent ultra-coarse-grained PC membrane model comprising three sites. Our goal is to parameterize the ultra-coarse-grained model to reproduce the experimentally observed structural and thermodynamic characteristics of PC membranes within real units and dimensions. This involves properties such as the area per lipid, area compressibility, bending modulus, line tension, phase transition temperature, density profile, and radial distribution function. To accomplish this, we conduct a series of evolutionary runs using diverse random initial population sizes, ranging from 96 to 384. As a result, our findings illustrate a direct correlation between increasing the initial population size and achieving lower cost values. Furthermore, we demonstrate that our ultra-coarse-grained membrane model exhibits realistic behaviors observed in lipid membranes. This includes self-assembly into bilayers, vesicle formation, membrane fusion, and gel phase formation. To enhance the accessibility of the ultra-coarse-grained membrane model for a broader audience, topology files of the derived potentials are supplied in a tabulated format. These files are meticulously designed to seamlessly integrate with the GROMACS molecular dynamics engine, facilitating immediate implementation and usage. Moreover, we highlight the compatibility of the developed ultra-coarse-grained model with the widely used Martini coarse-grained model. This compatibility enables the simulation of realistic membrane proteins within an ultra-coarse-grained bilayer membrane. This compatibility opens up avenues for researchers to explore complex membrane systems, delve into protein-membrane interactions, and gain valuable insights applicable to various biological processes and the development of therapeutics.

\begin{acknowledgement}
The authors would like to thank Marcus M\"uller for fruitful discussions. This work was funded by the Deutsche Forschungsgemeinschaft (DFG, German Research Foundation) under Germany's Excellence Strategy - EXC 2033 - 390677874 - RESOLV. 
\end{acknowledgement}

\bibliography{main.bib}
\end{document}